\documentclass[aip,jcp,amsmath,amssymb,floatfix,reprint,citeautoscript,noeprint,longbibliography]{revtex4-2}
\usepackage[utf8]{inputenc}
\usepackage{graphicx}
\usepackage{wrapfig}
\usepackage[colorlinks,allcolors=black,citecolor=blue,urlcolor=blue]{hyperref}
\usepackage[mathlines,displaymath]{lineno}
\usepackage{chemformula}
\usepackage{placeins}
\usepackage{booktabs}
\usepackage{parskip}
\DeclareUnicodeCharacter{2212}{-}
\graphicspath{{figures/}}

\usepackage{setspace}

\makeatletter
\def\@email#1#2{%
 \endgroup
 \patchcmd{\titleblock@produce}
  {\frontmatter@RRAPformat}
  {\frontmatter@RRAPformat{\produce@RRAP{#1\href{mailto:#2}{#2}}}\frontmatter@RRAPformat}
  {}{}
}%
\makeatother

\begin{document}

\title{The Wetting of \ch{H2O} by \ch{CO2}}


\author{Samuel G.\ H.\ Brookes}
\affiliation{%
Yusuf Hamied Department of Chemistry, University of Cambridge, Lensfield Road, Cambridge, CB2 1EW, UK
}
\affiliation{%
Cavendish Laboratory, Department of Physics, University of Cambridge, Cambridge, CB3 0HE, UK
}
\affiliation{%
Lennard-Jones Centre, University of Cambridge, Trinity Ln, Cambridge, CB2 1TN, UK
}
\author{Venkat Kapil}
\affiliation{%
Yusuf Hamied Department of Chemistry, University of Cambridge, Lensfield Road, Cambridge, CB2 1EW, UK
}
\affiliation{%
Lennard-Jones Centre, University of Cambridge, Trinity Ln, Cambridge, CB2 1TN, UK
}
\affiliation{%
Department of Physics and Astronomy, University College London, London, UK
}
\affiliation{%
Thomas Young Centre and London Centre for Nanotechnology, London, UK,  London, UK
}

\author{Christoph Schran}
\affiliation{%
Cavendish Laboratory, Department of Physics, University of Cambridge, Cambridge, CB3 0HE, UK
}
\affiliation{%
Lennard-Jones Centre, University of Cambridge, Trinity Ln, Cambridge, CB2 1TN, UK
}

\author{Angelos Michaelides}\affiliation{%
Yusuf Hamied Department of Chemistry, University of Cambridge, Lensfield Road, Cambridge, CB2 1EW, UK
}
\affiliation{%
Lennard-Jones Centre, University of Cambridge, Trinity Ln, Cambridge, CB2 1TN, UK
}
\email{vk380@cam.ac.uk  \: cs2121@cam.ac.uk  \:  am452@cam.ac.uk}
\date{June 2024}

\begin{abstract}
Biphasic interfaces are complex but fascinating regimes that display a number of properties distinct from those of the bulk.
The \ch{CO2}-\ch{H2O} interface, in particular, has been the subject of a number of studies on account of its importance for the carbon life cycle as well as carbon capture and sequestration schemes.
Despite this attention, there remain a number of open questions on the nature of the \ch{CO2}-\ch{H2O} interface, particularly concerning the interfacial tension and phase behavior of \ch{CO2} at the interface.
In this paper, we seek to address these ambiguities using \textit{ab initio}-quality simulations. 
Harnessing the benefits of machine-learned potentials and enhanced statistical sampling methods, we present an \textit{ab initio}-level description of the \ch{CO2}-\ch{H2O} interface.
Interfacial tensions are predicted from 1-500 bar and found to be in close agreement with experiment at the pressures for which experimental data is available. 
Structural analyses indicate the build-up of an adsorbed, saturated \ch{CO2} film forming at low pressure (20 bar) with properties similar to those of the bulk liquid, but preferential perpendicular alignment with respect to the interface. 
\ch{CO2} monolayer build-up coincides with a reduced structuring of water molecules close to the interface.
This study highlights the predictive nature of machine-learned potentials for complex macroscopic properties of biphasic interfaces,
and the mechanistic insight obtained into carbon dioxide aggregation at the water interface is of high relevance for geoscience, climate research, and materials science.
\end{abstract}

\maketitle

\section{Introduction}
The boundary between two immiscible fluids represents a unique and fascinating regime. 
The properties of this biphasic interface, noticeably different from those of bulk fluid, arise from its inhomogeneous composition and the anisotropic distribution of its constituent particles \cite{Benjamin1993}.
The resulting asymmetry in forces acting between these particles promotes a series of interesting phenomena, including molecular orientational aligment, density gradients and structuring, phase transfer effects, and variations in reactivity and dielectric properties \cite{Benjamin1993,Ruiz-Lopez2020,Shah2020,Gonella2021}.

Arguably the most important biphasic interface is that of carbon dioxide and water, \ch{CO2}-\ch{H2O}. 
This interface exhibits a diverse range of physiochemical properties, which stem from the distinct phases that \ch{CO2} can form with subtle changes to temperature and pressure.
For ambient temperatures, \ch{CO2}-\ch{H2O} exists as a gas-liquid interface for pressures $p$ less than the critical pressure ($p_{\mathrm{C}}$ = 73.8 bar, $T_{\mathrm{C}}$ = 304.1 K) \cite{Span1996}. 
At higher pressures, \ch{CO2}-\ch{H2O} forms a liquid-liquid (or supercritical-liquid) system, suggesting that temperature or pressure can be used as a lever to modulate biphasic properties. 
\ch{CO2}-\ch{H2O} finds use in a variety of real-word applications and phenomena, from compressible solvent design \cite{Medina-GonzalezYaocihuatl2014SSBP} to ocean acidification monitoring \cite{HONISCH2005305} to enhanced oil and gas recovery schemes \cite{Klewiah2020}.
They are also central to various carbon capture and storage (CCS) techniques, where anthropogenic \ch{CO2} is captured at source and injected under pressure into predominately aqueous subsurface storage sites, e.g., saline aquifers or disused coal seams \cite{Leung2014}.
In these storage sites, retention of the anthropogenic carbon is mediated by the interactions between injected \ch{CO2} and \textit{in situ} \ch{H2O}, meaning that both storage capacity and the duration are functions of the interfacial behavior.

\begin{figure*}[t]
    \centering
    \includegraphics[scale=0.525]{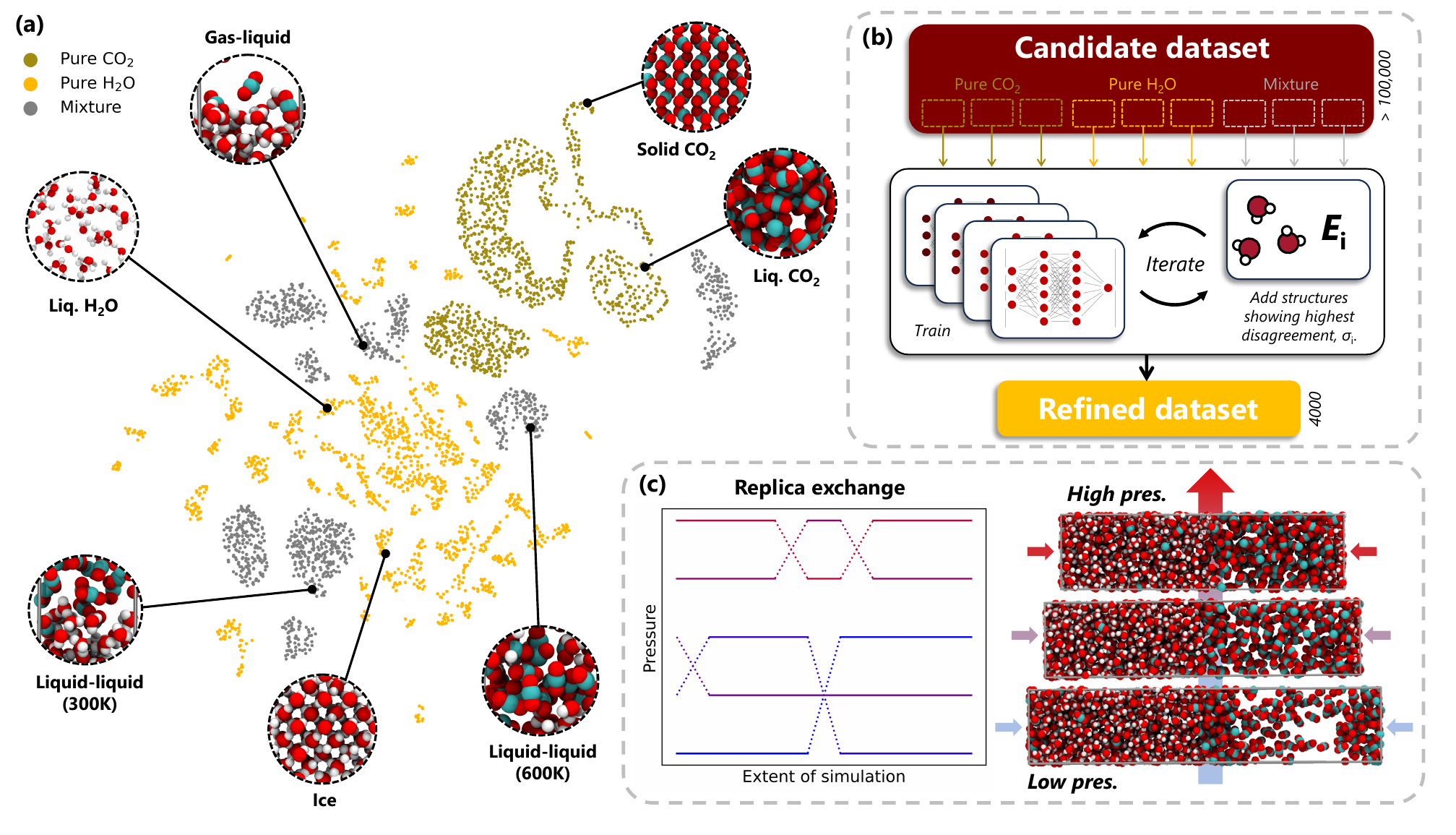}
    \caption{\label{fig:overview}
    Modeling biphasic fluid interfaces.
    \textbf{(a)} 2D projection of the structural data used to train our neural network potential.
    \textbf{(b)} Query-by-committee workflow for the creation of a compact, structurally diverse dataset for each constituent trajectory. 
    \textbf{(c)} Application of the model using REMD towards the determination of macroscopic and microscopic properties.
    }
\end{figure*}

Realizing these applications for \ch{CO2}-\ch{H2O} requires a detailed knowledge of its biphasic interface. 
This includes understanding both its microscopic properties (in terms of the structuring of molecules, their orientations, the nature of their bonding, etc.) as well as its macroscopic properties, such as phase densities and interfacial tensions (IFTs), $\gamma$.
The IFT, in particular, is critical to understanding biphasic interfaces.
This property encapsulates much of the macroscopic nature of the interfacial region and is used to gauge their relative miscibilities. 
In terms of CCS and \ch{CO2}-\ch{H2O}, $\gamma$ is directly proportional to the sealing capillary pressure, which quantifies the volume of \ch{CO2} than can be stored and thus is a good estimate of the efficacy of a chosen storage site \cite{Nielsen2012,Hossein2022}.

An extensive picture of both micro- and macroscopic properties of \ch{CO2}-\ch{H2O} has been built up over the last decades. 
In experimental literature, researchers have focused mostly on the measurement of $\gamma$ under differing temperatures and pressures \cite{Chun1995,Hebach2002,Tewes2004,Chiquet2007,Bachu2009,Georgiadis2010,Aggelopoulos2010,Aggelopoulos2011,Bikkina2011,Liu2017,Hinton2021}. 
In general, $\gamma$ is observed to decrease with either increasing pressure or increasing temperature, though the exact nature of this reduction is dependent upon which CO$_2$ phase is present (see later, Fig.\ \ref{fig:ift}). 
Measurements from experiment, however, are subject to significant variation depending on experimental setup (e.g., thermocouple placement, equilibration times, density treatment) \cite{Nielsen2012}. 
Therefore, experimental work is often supplemented by theoretical and computational estimates for $\gamma$.
Statistical perturbation models, such as statistical associating fluid theory (SAFT), do a good job in reproducing experimental values but can sometimes lack sufficient microscopic insight on the origins of these results \cite{Xiao-Sen2008,NA2012,Lafitte2010}. 
In contrast, force-field (FF) MD simulations provide both estimates for $\gamma$ and full atomistic detail on the behavior of \ch{CO2}-\ch{H2O}.
Although the exact description of the \ch{CO2}-\ch{H2O} interface is sensitive to the choice of the empirical parameters, FFs have provided several important trends and insights. 
These include: \ch{CO2}-\ch{H2O} interfaces are molecularly sharp; molecules form a layered structure within several angstrom of the interface; water's dipole orients parallel to the interface; and there exists a strong, lateral network of hydrogen bonds within the first contact layer of water \cite{DaRocha2001,Zhang2011,Nielsen2012,Zhao2015,Li2020,Shiga2023}.

Despite the valuable insight uncovered from these studies, a number of open questions remain surrounding \ch{CO2}-\ch{H2O}.
For example, how exactly does $\gamma$ behave in the vicinity of \ch{CO2}'s critical point ($p_{\mathrm{C}}$, $T_{\mathrm{C}}$)?
How do high pressures impact the value of $\gamma$? 
And what is the nature of interfacial \ch{CO2} and \ch{H2O} molecules and how exactly do they interact with one another?

Building on previous work in the field, this paper aims to provide an new perspective for characterizing the \ch{CO2}-\ch{H2O} interface. 
Our work focuses on generating machine-learned potentials (MLPs), which we use to perform MD simulations with an \textit{ab initio} level of accuracy.
Unlike previously applied methods, our work breaks away from the empirical fitting to experimental or literature values and instead provides estimates from the ground up. 
MLPs already have a proven history of success in simulating complex systems~\cite{Unke2019,Deringer2019,Kapil2022,YANG2022143,Litman2023,Mathur2023,omranpour2024perspective}, and their use in simulating \ch{CO2}-\ch{H2O} would help address many of the open questions on this system. 
To date, however, an \textit{ab initio} or MLP description of biphasic fluid interfaces has remained elusive.
Biphasic systems by definition require larger simulation boxes than bulk for direct \textit{ab initio} simulations and for generating training data for MLPs. 
At the same time, converging the properties of the bulk and interfacial regions requires long simulation time, beyond the reach of standard molecular dynamics.

In this work, we bring together \textit{ab inito} accuracy and thorough sampling to accurately simulate large and complex mixtures. 
We utilize a number of distinct methodological features - specifically, MLPs for fast and accurate \textit{ab initio} total energy predictions, active learning strategy to build a compact and representative training sets, and replica exchange simulations for enhanced sampling of the interface
Our approach allows accurate and converged analysis of both macro- and microscopic properties of the \ch{CO2}-\ch{H2O} interface.
We obtain estimates of the interfacial tension at room temperature as a function of pressure.
These are found to be in good agreement with literature and help discern experimental reference data across the gas-liquid phase transition for \ch{CO2}. 
Having obtained a correct description of the interfacial tension, we study the structural and dynamical properties of the \ch{CO2}-\ch{H2O} interface to obtain key microscopic insights.
Notable observations include the buildup of a liquid-like CO$_2$ for low pressures and a reduced structuring in the aqueous phase with increasing pressure. 
We anticipate the future application of our potential to other important state points for \ch{CO2}-\ch{H2O} and that our approach will serve as a robust blueprint for investigating other complex biphasic interfaces.

\section{Methods}
Our approach to modeling the biphasic \ch{CO2}-\ch{H2O} interface comprises three established components: high-dimensional neural network potentials~\cite{PhysRevLett.98.146401,doi:10.1063/5.0016004,doi:10.1073/pnas.2110077118} (HD-NNPs) to act as our first-principles force generator; Query by Committee (QbC), an active learning technique for generating and optimising our structural dataset \cite{doi:10.1073/pnas.2110077118}; and replica exchange molecular dynamics (REMD) to expedite the statistical convergence of interfacial properties over multiple thermodynamic state points. 
Combining these features allows us to apply accurate potential energy surfaces for simulating large, complex molecular systems over extended time periods. 
An overview of our approach is provided in Fig.\ \ref{fig:overview}.

\subsection*{Training}

Our work follows the committee NNP development procedure outlined in Ref.~\citenum{doi:10.1073/pnas.2110077118}. 
We employ Behler-Parrinello HD-NNPs trained on DFT-level \textit{ab initio} total energies and forces estimated with CP2K \cite{PhysRevLett.98.146401,doi:10.1063/5.0007045}. 
A set of hand-crafted radial and angular symmetry functions were used \cite{doi:10.1073/pnas.2110077118}.
We employed the BLYP functional \cite{B1988,LYP1988} augmented by Grimme’s D3 corrections \cite{doi:10.1063/1.3382344}.
This setup has been shown to closely reproduce the condensed-phase and structural properties of both pure \ch{CO2} and pure \ch{H2O} \cite{Gillan2016,Goel2018,Mathur2023}.
Additionally, BLYP-D3 replicates the critical and vapor-liquid coexistence properties for \ch{CO2}, making it an appropriate selection for treating the various \ch{CO2} phases exhibited within our chosen pressure range \cite{Goel2018,Mathur2023}. 
Goedecker-Teter-Hutter (GTH) pseudopotentials were used for treatment of the core electrons, TZV2P-GTH basis sets for the valence electron density, and a planewave cutoff of 1050\,Ry was used.

Training of our NNPs was implemented using QbC over multiple improvement cycles (see Fig.~\ref{fig:overview}b for a schematic description). 
This was performed using the open-source AML package \cite{doi:10.1073/pnas.2110077118}. 
A candidate dataset was provided in the form of chemical structures labeled with energies and atomistic forces. 
Structures are generated using a mixture of flexible force field potentials (SPC/Fw \cite{10.1063/1.2136877} + Zhu \cite{ZHU2009}, Lorentz-Berthelot combining rules) as well as preliminary NNPs across three active learning cycles. 
This gave a total of more than 100,000 structures. 
QbC was used to trim and optimize this dataset through the iterative selection of those structures displaying the largest uncertainties across each committee member. 
This was repeated for each trajectory, resulting in a refined dataset of some 4000 structures.
Using this dataset, a final model was trained using the openly available N2P2 package \cite{Singraber2019}.

This final, refined dataset is shown in Fig.\ \ref{fig:overview}a.
A breadth of structures is included here - including gas-liquid, liquid-liquid, crystalline, and pure-phase structures - in order to maximize exploration of the relevant configuration space. 
A committee of NNPs was trained using N2P2\cite{Singraber2019}, and the committee member displaying the lowest errors was selected as the final model. 
This final NNP was validated against the predictions of \textit{ab initio} MD in the form of energies and forces (test RMSEs: 3.31 meV/atom and 88.26 meV/\AA{}, of similar performance to that reported in Ref.\ \citenum{Morawietz2016} for pure water) as well as structural predictions of pure-phase properties, which are detailed in Section 1 of the SI.

\subsection*{Simulation setup}

Typical system setups for our production run are shown in Fig.\ \ref{fig:overview}c.
Each system consisted of 600 water molecules and 200 \ch{CO2} molecules.
Lateral dimensions $xy$ were fixed at 20 \AA{}, and periodic boundary conditions were applied along all three axes. 
Atomic configurations were initialized such that the \ch{CO2}-\ch{H2O} interface resided parallel to the $xy$ plane. 
System sizes and compositions were chosen to mitigate periodic error and finite-size effects, which have been shown in FF-MD to impact IFTs and other interfacial properties \cite{Janecek2009,Longford2018}. 
Results from our testing of IFT against system size are shown in Section 2 of the SI.

\subsection*{Replica exchange simulations}

15 representative pressures were selected spanning the 1-500 bar range. 
REMD was performed across 15 different pressures, from which we extracted macropscopic measurements (i.e., IFTs) as well as microscopic properties in the form of density profiles, molecular orientations, and radial distribution functions (RDFs).
NNP-REMD was performed using the open-source \textit{i-pi} package \cite{Kapil2019} with \textit{LAMMPS} as the force generator \cite{LAMMPS,Singraber2019a}. 
For each state point, simulations were run for 10\,ns of simulation time corresponding to $10^7$ steps. 
These were performed under the $NP_{\mathrm{z}}AT$ ensemble, with fixed number of particles $N$, fixed lateral ($xy$) area $A$, fixed temperature (300 K), and $P_{\mathrm{z}}$ set to the target pressure. 
Collectively, these simulations compile 150\,ns of \textit{ab initio}-quality data, forming a robust basis for our analysis.
We have checked the sensitivity of our results on the choice of DFT functional by also training an NNP model to revPBE-D3.
Resulting IFTs prove to be very similar overall, and these are shown in Section 3 of the SI.

\begin{figure}[t]
    \centering
    \includegraphics[scale=0.555]{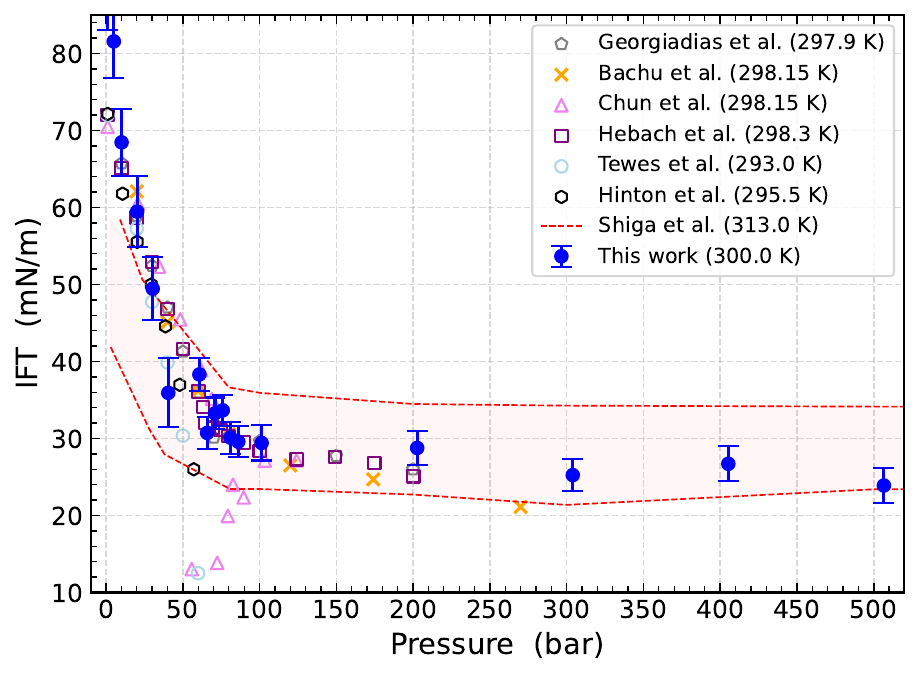}
    \caption{\label{fig:ift}
    Measuring the interfacial tension ($\gamma$) of \ch{CO2}-\ch{H2O}.
    Interfacial tension results for our neural network potential (blue, filled) measured at 300\,K for pressures 1-500\,bar.
    Error bars denote 2$\sigma$, where $\sigma$ is the standard error calculated from block averaging. 
    Selected previous experimental results are shown as hollow markers, whilst the computational work of Shiga et al.\ (FF-MD) is encompassed within the shaded region \cite{Shiga2023}.
    }
\end{figure}

\section{Results}
\subsection*{IFT profile replicates experimental values}
The interfacial tension is a key property for determining the miscibility of two phases. 
A number of previous experimental and computational studies have sought to characterize the IFT of \ch{CO2}-\ch{H2O} as a function of pressure. 
In Fig.~\ref{fig:ift}, we plot the results of these studies at roughly room temperature alongside our own estimates obtained using the developed NNP.
We compute $\gamma$ using the statistical mechanical route of Kirkwood and Buff (KB) \cite{Kirkwood1951}, 
\begin{equation}
\label{eq:ift}
    \gamma = \frac{L_{\mathrm{z}}}{2} \left(P_{\mathrm{||}} - P_{\mathrm{\perp}}  \right)
\end{equation}
where $L_{\mathrm{z}}$ gives the $z$ length of the simulation box, and $P_{\mathrm{||}}$ and $P_{\mathrm{\perp}}$ represent the normal and lateral pressure tensor components, respectively.

\begin{figure}[t]
    \centering
    \includegraphics[scale=0.57]{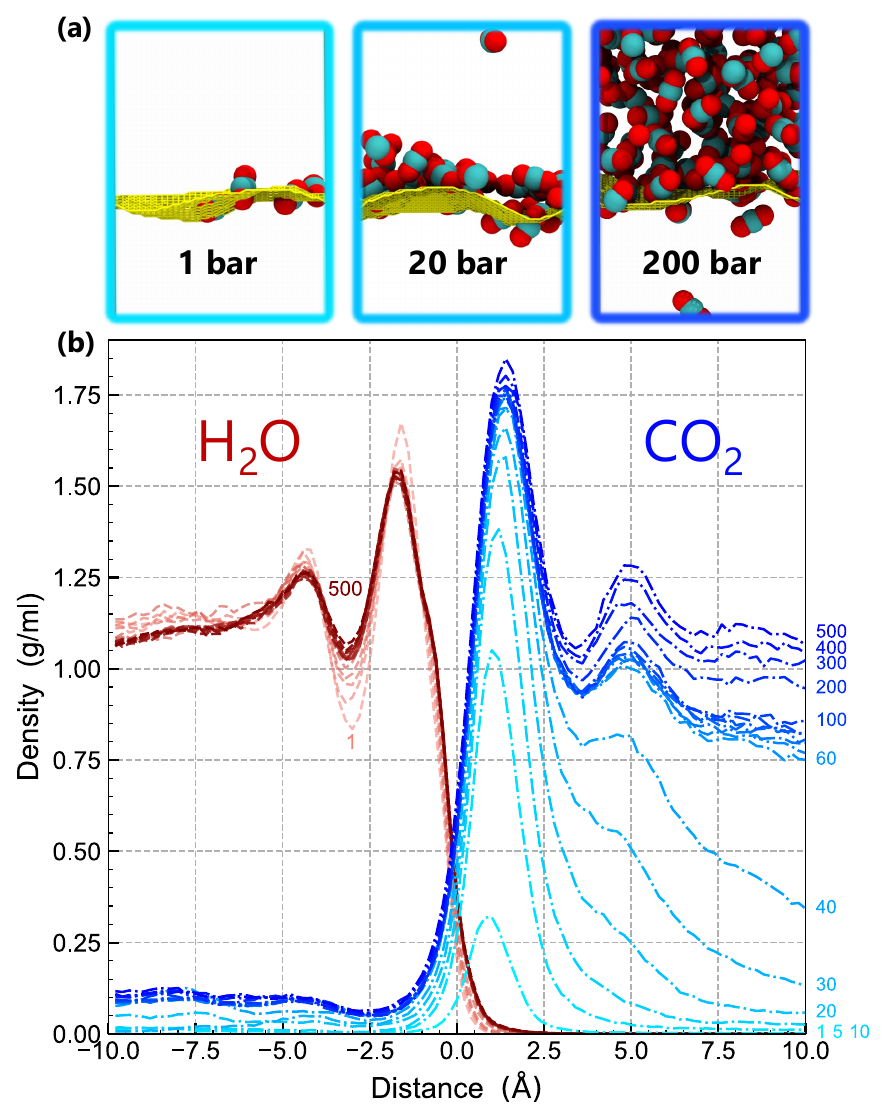}
    \caption{\label{fig:density}
    Profiling the change in \ch{CO2}-\ch{H2O} from 1 to 500\,bar.
    \textbf{(a)} Representative snapshots of the \ch{CO2} phase and instantaneous interface (yellow mesh) for 1, 20, and 200\,bar. \ch{H2O} has been omitted for visual clarity. 
    \textbf{(b)} Density profiles are shown at different pressures for water (red) and \ch{CO2} (blue) with distance from the instantaneous interface that separates them.
    }
\end{figure}

Inspection of Fig.~\ref{fig:ift} yields several observations.
For low pressures ($p \: <$ 60\,bar), our simulations predict a steep descent in $\gamma$ with increasing pressure.
This phenomenon is associated with the gradual accumulation of \ch{CO2} at the water surface, a process which reduces the spatial anisotropy at the phase boundary and thus also the magnitude of $\gamma$ (see later, Fig. \ref{fig:density}). 
For intermediate pressures (60 $\leq \: p \: <$ 100\,bar), our NNP results indicate an abrupt change in the behavior of $\gamma$ as we cross the critical pressure boundary (73.8 bar), going from gas-liquid to liquid-liquid. 
And finally for high pressures ($p \: \geq $ 100\,bar), our model yields a slight reduction in $\gamma$ with increasing pressure. 
The comparatively small gradient in $\gamma$ for this regime is attributed to a saturation in the concentration of \ch{CO2} close to the interface, such that changes to the pressure lead to minimal change in the near-surface environment. 
A minimum of 24 $\pm$ 2 mN/m is recorded at 500 bar.

Across Fig.~\ref{fig:density}, we secondly observe close agreement between experimental and our simulation results, within the range of pressures for which experimental data is available. 
This is true for both low pressures ($p \: <$ 60\,bar), where ${\gamma}_{\mathrm{NNP}}$ replicates the steep descent in gas-liquid IFTs, and high pressure ($p \: \geq$ 100\,bar), where ${\gamma}_{\mathrm{NNP}}$ follows the gradual decline in liquid-liquid IFTs as reported by Hebach et al.\ \cite{Hebach2002} and Bachu et al.\ \cite{Bachu2009} 
A notable aspect of this work is the extension of IFT measurements beyond the currently available experimental pressure range for $\sim$ 300 K. 
Moreover, ${\gamma}_{\mathrm{NNP}}$ values recorded for intermediate pressures (60 $\leq \: p \: <$ 100\,bar) provide additional clarity on the nature of $\gamma$ in the vicinity of CO$_2$'s phase transition pressure $p_{\mathrm{T}}$ (73.8\,bar). 
This region has previously been a point of contention, with experimental results differing by up to 20 mN/m for pressures close to $p_{\mathrm{T}}$, as seen by the selection of results displayed in Fig.~\ref{fig:ift}.
Our results clearly support the presence of a break in the slope between gas-liquid and liquid-liquid regimes, as reported by a number of authors \cite{Georgiadis2010,Bachu2009,Hebach2002}. 
There is no evidence of the `dip' or `cusp' in $\gamma$ near $p_{\mathrm{T}}$ as predicted by some experiments \cite{Chun1995,Tewes2004}.
Such observations are now thought to be experimental artifacts that stem either from differences in thermocouple placement or the inherent uncertainty in near-phase-transition properties \cite{Hebach2002,Nielsen2012}.

Our results also qualitatively corroborate the results of previous FF simulations. 
These are encompassed within the shaded portion of Fig.\ \ref{fig:ift}, the limits of which represent the extent of $\gamma$ predictions for varying combinations of FF models (e.g., SPC/E + EPM2, SPC/E + PPL, TIP4P/2005 + PPL, etc.) \cite{Shiga2023}. 
Whilst the exact value of $\gamma$ is sensitive to the choice of models and parameterization, the overall range of FF predictions is in qualitative agreement with our NNP predictions, with a notable exception of the low pressure regime \cite{Shiga2023}.
For these pressures, classical FFs systematically underpredict $\gamma$ by 15-30 mN/m. 
We attribute differences in the performance of FF and NNP models to the differing levels of underlying theory and the fact that our NNP is trained explicitly to replicate the interactions between \ch{CO2} and \ch{H2O}.
In comparison, bicomponent FFs are constructed using geometric/mathemical mixing rules, the choice of which is often \textit{ad hoc} and can lead to substantial variation in the resulting thermodynamic properties \cite{DELHOMMELLE2001,Boda2008,Rouha2009}.
This is exemplified by the large spread in $\gamma$ values shown for FF-MD in Fig.\ \ref{fig:ift}.

\subsection*{\ch{CO2} forms a saturated monolayer at low pressure}

To better understand the interfacial tension shown in Fig.\ \ref{fig:ift}, we plot a series of microscopic profiles detailing the variation in density $\rho$ with distance $d$ from the instantaneous interface. 
This interface is calculated at each timestep using the Willard-Chandler formalism \cite{Willard2010}, allowing for a full resolution of the dynamic nature of the interface. 
Resulting profiles are plotted in Fig.~\ref{fig:density} for both \ch{H2O} (red) and \ch{CO2} (blue). An inspection of Fig.~\ref{fig:density} reveals a stark contrast in the phase behavior of water and \ch{CO2}. 
Focusing first on the water phase, we observe a layered structuring in $\rho$ across all pressures, in close agreement with the behavior of the air-water interface~\cite{Litman2023}. 
The shape of these profiles is relatively unvarying with changes to the pressure, though we do note an enhanced structuring of molecules for $p \: \leq$ 20\,bar approaching atmospheric pressure. 
In contrast to this behavior, the \ch{CO2} profile exhibits large changes in both magnitude and shape with changes to the pressure. 
For $p \: <$ 10\,bar, density profiles register a sharp peak at roughly 1\,\AA{} with a density that tails off exponentially with distance from the interface. 
Physically, this represents a film of monolayer \ch{CO2} adsorbed at the water surface and a gaseous CO$_2$ phase extending beyond to large distance. 
For 20 $\leq \: p \: < $ 60\,bar, a second peak is observed at 4\,\AA{}, suggesting the formation of a bilayer of \ch{CO2} at the interface. 
Beyond 60\,bar, the exponential decay in density is replaced by liquid-like layering that tends towards bulk density for large distances. 
Further increases in pressure beyond 100\,bar serve to increase both peak heights and bulk density as $d \: \rightarrow \: \infty$.

Relating these microscopic observations with the IFT profile of Fig.~\ref{fig:ift}, it is clear that the variation in $\gamma$ is a product of the considerable changes observed in \ch{CO2} with changing pressure. 
Unlike water, \ch{CO2} is a nonpolar molecule that exhibits relatively weak intermolecular interactions. 
Its critical pressure $p_{\mathrm{C}}$ resides within our range of sampling, so we end up simulating two distinct phases of \ch{CO2} as well as a phase transition region. 
In comparison, at 300\,K, water is firmly within the liquid phase, and its high natural surface tension - a product of its extensive hydrogen bonding network - ensures preservation of its structural integrity and prevents extensive mixing with the \ch{CO2} phase. 
The most significant change in the structuring of water occurs for low pressure ($p \: \leq$ 20 bar). 
It is interesting to note that this change coincides with the emergence of the monolayer in \ch{CO2}. 
The combination of these observations possibly suggests that structuring of the aqueous phase is mediated by the extent of monolayer coverage of the adsorbed phase and that subsequent adsorbed layers have minimal impact.

Our \textit{ab initio}-level profiles share many similarities with previous force field modeling \cite{DaRocha2001,Zhang2011}. 
This includes the formation of monolayer and bilayer \ch{CO2} and the recovery of liquid-like properties at high pressure. 
Interestingly, however, the pressure at which our model predicts the emergence of a \ch{CO2} monolayer (and therefore also the starting of a second \ch{CO2} layer) is much lower compared to conventional FF predictions (e.g., those of SPC/E + EPM2). 
In our results, a fully saturated monolayer forms at 20\,bar; in FFMD studies, this occurs at higher pressures (around 60\,bar for SPC/E+EPM2), thereby suggesting a lower wettability for \ch{CO2} \cite{Zhang2011}. 
We again attribute differences in the predictions of these models to the underlying theory, with NNPs trained to account for both polarization and charge transfer effects. 
On the other hand, the classical FFs used so far to study this interface are limited by the nature of their mixing rules and their rigid-body formulation, which we would not expect to fully represent the interfacial regime.

\subsection*{Monolayer shows liquid-like properties}
\begin{figure}[t]
    \centering
    \includegraphics[scale=0.62]{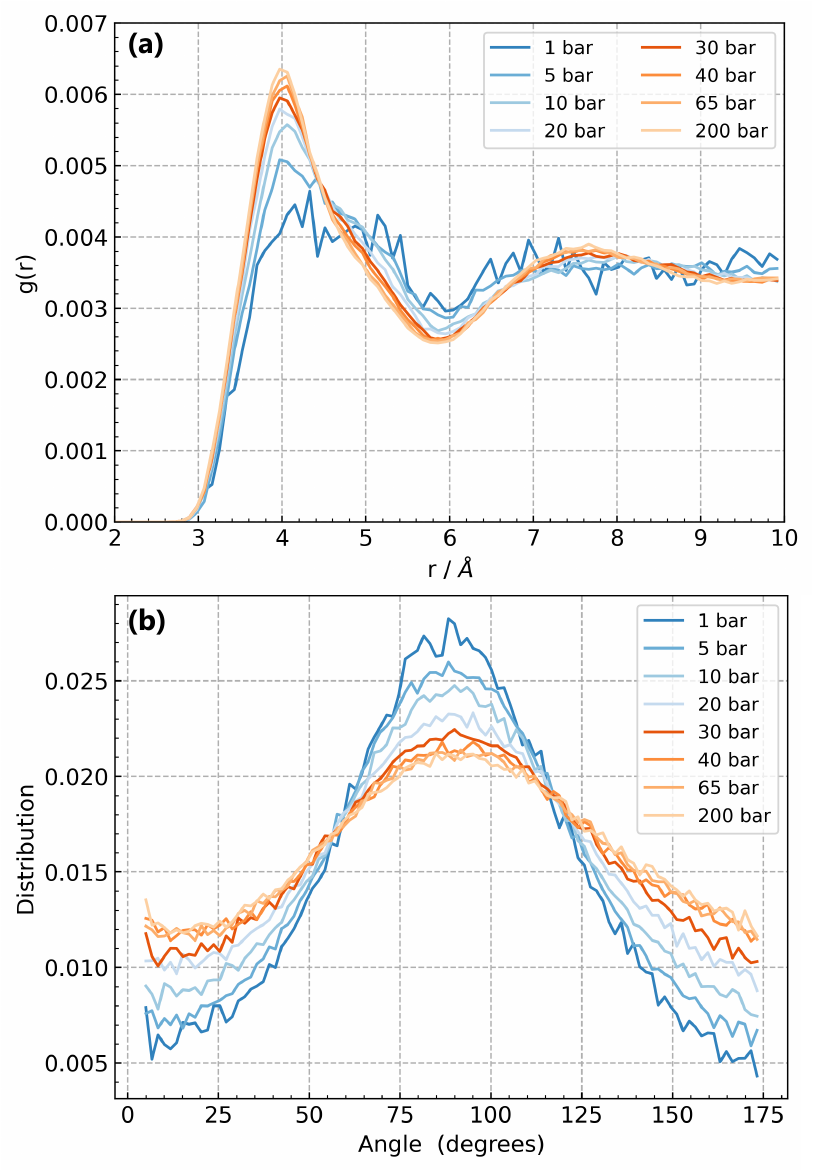}
    \caption{\label{fig:co2-layer}
    Characterising \ch{CO2} build-up. 
    \textbf{(a)} LDFs for \ch{CO2} located within first layer.
    \textbf{(b)} Angular distributions of \ch{CO2} located within first layer.
    Angles are calculated between C-O vector, $\mathrm{\boldsymbol{v}_{CO}}$, and the surface-molecule vector, $\mathrm{\boldsymbol{v}_{surf}}$.
    The latter is defined as being the vector connecting the carbon of a \ch{CO2} molecule and the point on the instantaneous surface closest to that molecule.
    }
\end{figure}

Results from the microscopic profiling of \ch{CO2}-\ch{H2O} show the formation of a saturated \ch{CO2} monolayer at low pressure.
To better understand this phenomenon, we investigate the structural properties of \ch{CO2} within 2.5\,\AA{} of the instantaneous interface, i.e., within the first contact layer. 
Fig.~\ref{fig:co2-layer} plots both the 2D lateral distribution function (LDF) and angular distribution of these molecules. 
The LDF, $g(r)$ is calculated such that it accounts for the quasi-2D nature of this monolayer
\begin{equation}
\label{eq:rdf}
    g(r) = \frac{dn_{\mathrm{r}}}{dV \cdot \rho} = \frac{dn_{\mathrm{r}}}{2 \pi dr \cdot \rho}
\end{equation}
where $dn_{\mathrm{r}}$ is the number of \ch{CO2} molecules within a shell of thickness $dr$, and $\rho$ gives the local density.
Inspection of Fig.\ 4a shows how the structuring of this layer quickly converges towards a bulk-like character.
By 20\,bar, we see a distribution that emulates that of bulk \ch{CO2} and suggests a liquid-like nature for the first contact layer.
This is true for pressures 20-500\,bar.

In Fig.~\ref{fig:co2-layer}b, we show the distribution of angles between the instantaneous interface and the \ch{CO2} orientation. 
Across all pressures, our results suggest a clear preference for \ch{CO2} to lie flat at the water surface. 
This corroborates previous computational work, which suggest that interactions between water and \ch{CO2} are maximized when the latter molecule lies parallel at the surface \cite{DaRocha2001}. 
We note that the distribution is more pronounced at 90 degrees for $p \: <$ 20\,bar, where the adsorbed CO$_2$ is more exposed and has fewer molecules to interact with. 
The addition of more adsorbed molecules at the aqueous surface reduces the strength of water-\ch{CO2} interactions, allowing for greater freedom in the latter molecule's orientational alignment with increasing pressure.

\section{Summary and Outlook}
Biphasic interfaces represent complex and dynamic regimes. 
In this paper, we have highlighted an approach for analyzing these regions in a way that combines \textit{ab initio} accuracy with converged statistics and computational tractability. 
Our methodology allowed for nanosecond treatment of large \ch{CO2}-\ch{H2O} systems, which has yielded a converged IFT profile at an \textit{ab initio} level of accuracy. 
Our results show good agreement with experimental literature for low- and high-pressure regimes, and microscopic insight into this behavior has been provided. 
The reproduction of results in the vicinity of a phase transition (gas to liquid) is notable given the difficulty associated with treating this highly dynamic regime. 
We observe the formation of a saturated \ch{CO2} monolayer at low pressure (20 bar) with structural properties akin to those of bulk \ch{CO2}. 
The emergence of this monolayer coincides with reduced structuring for near-interface water molecules, suggesting that interactions between the first contact layers of water and \ch{CO2} are critical for aqueous phase structuring. 
We envisage that such insights will be important for realizing \ch{CO2}-\ch{H2O}'s many applications as well as shedding light on other significant biphasic systems, for example, in biological membranes \cite{Pattni2015a} and for liquid-liquid interfaces facilitating nanoparticle assembly \cite{dasgupta2017nano,Sokoowski2019}

In terms of carbon capture and sequestration, our NNP provides a robust tool for extending IFT measurements and providing benchmark figures for higher pressure regimes. 
In this way, coupled with additional geological measurements, one might use our model to provide estimates for a particular storage site of known temperature and pressure, thereby also providing an estimate of that site's suitability.  
Compared with statistical perturbation models, our work has the advantage of providing additional microscopic insight to supplement predictions of $\gamma$.
Future work could look to exploit this microscopic insight towards providing estimates on the contact angle between \ch{CO2} and \ch{H2O}, another property of import for gauging storage efficacy. 
In addition, we could extend our models to incorporate explicit ions (e.g., Na$^+$ and Cl$^-$) to better replicate the saline conditions of underground storage. 
Understanding how these ions behave and impact the IFT profile will add a further layer of realism onto this work and allow for better estimations of storage lifetimes and capacities.

Our work also unearths several differences between classical and \textit{ab initio}-level modeling of the \ch{CO2}-\ch{H2O} interface. 
Comparing with previous classical results, we suggest that previous analyses of low-pressure \ch{CO2} coverage in aqueous systems may require reanalyzing. 
This could have profound implications for our understanding of process such as \ch{CO2} adsorption and its role in ocean acidification. 
Higher levels of \ch{CO2} adsorption at the ocean surface imply greater \ch{CO2} dissolution and therefore higher overall pH. 
Extending our current model through the addition of other relevant species, e.g., carbonate, bicarbonate, and carbonic acid, it would be interesting to investigate this acidification process and understand how the adsorption, dissolution, and subsequent reaction of \ch{CO2} proceeds under differing conditions.

In conclusion, we provide new insight on the nature of biphasic interfaces and demonstrate the significance of interactions occurring between molecules adsorbed in the first contact layer. 
We anticipate application of the combined set of techniques used here to other important biphasic systems.
The immediate findings of this study about preferred \ch{CO2} layering at \ch{H2O} are expected to be of direct relevance in geoscience, climate research, and material science.

\section*{Supplementary Material}
\noindent See the supplementary material for NNP validation tests, system-size convergence tests, and a comparison of IFT predictions for NNP and classial MD.

\section*{Acknowledgments}
SGHB is supported by the Syntech CDT and funded by EPSRC (Grant No.\ EP/S024220/1).
CS acknowledges financial support from the Deutsche Forschungsgemeinschaft (DFG, German Research Foundation) project number 500244608.
V.K. acknowledges support from the Ernest Oppenheimer Early Career Fellowship and the Sydney Harvey Junior Research Fellowship, Churchill College, University of Cambridge. 
AM acknowledges support from the European Union under the “n-AQUA” European Research Council project (Grant No. 101071937).
We are grateful for computational support and resources from the UK Materials and Molecular Modeling Hub which is partially funded by EPSRC (Grant Nos. EP/P020194/1 and EP/T022213/1).
We are also grateful for computational support and resources from the UK national high-performance computing service, Advanced Research Computing High End Resource (ARCHER2) and the Swiss National Supercomputing Centre under project s1209.
Access for both the UK Materials and Molecular Modeling Hub and ARCHER2 were obtained via the UK Car-Parrinello consortium, funded by EPSRC grant reference EP/P022561/1.
Access to CSD3 was obtained through a University of Cambridge EPSRC Core Equipment Award (EP/X034712/1).

\section*{Author Declarations}
\subsection*{Conflict of Interest}
\noindent The authors have no conflicts to disclose

\subsection*{Author Contributions}
\noindent \textbf{Samuel G.\ H.\ Brookes:} Conceptualization (equal); Investigation (lead); Formal analysis (equal); Writing - original draft (Lead); Writing - review and editing (equal).
\textbf{Venkat Kapil:} Conceptualization (equal); Methodology (equal); Formal analysis (equal); Writing - review and editing (equal); Supervision (equal). 
\textbf{Christoph Schran:} Conceptualization (equal); Methodology (equal); Formal analysis (equal); Writing - review and editing (equal); Supervision (equal). 
\textbf{Angelos Michaelides:} Conceptualization (equal); Formal analysis (equal); Writing - review and editing (equal); Supervision (equal).

\section*{Data Availability}
\noindent All data required to reproduce the findings of this study are available at XXX. Data will be made available upon acceptance of the paper.

\

\bibliography{MAIN}

\end{document}


\maketitle

\tableofcontents

\section{Model validation}

\begin{figure*}[h]
    \centering
    \includegraphics[scale=0.7]{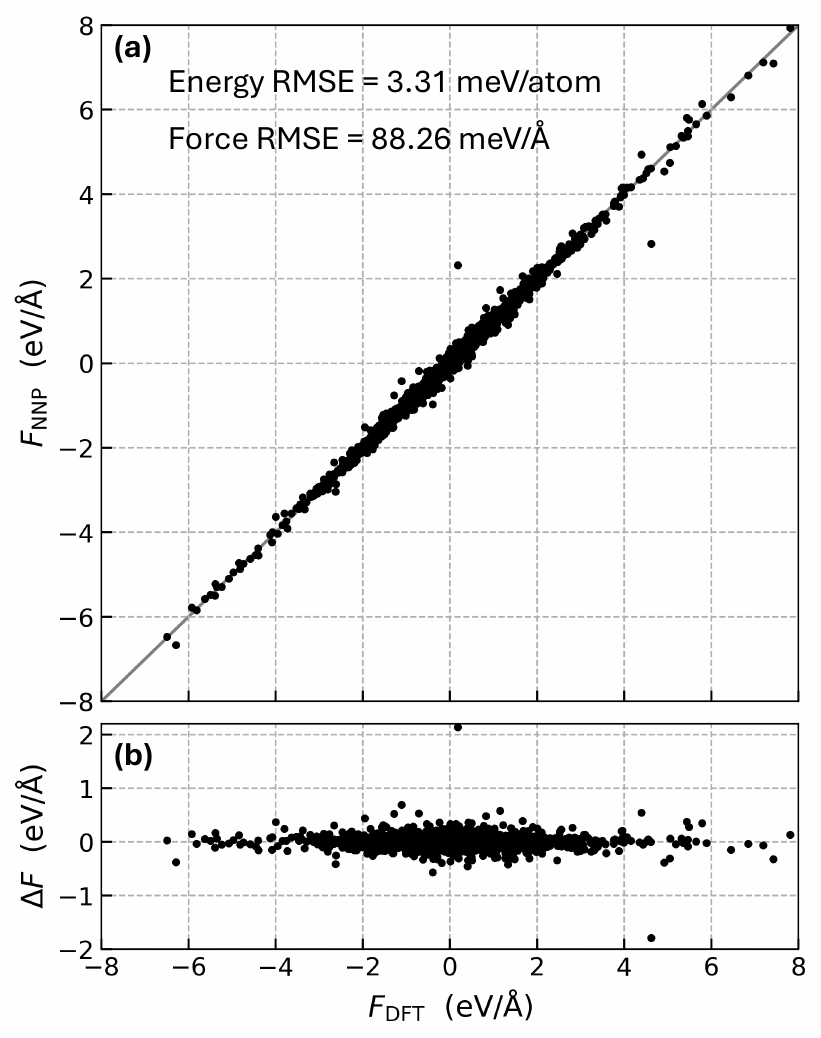}
    \caption{\label{fig:rmse}
    Comparing NNP and reference DFT force predictions. 
    \textbf{(a)} Correlation plot showing NNP forces against DFT (BLYP-D3) forces for a test set of 400 structures. 
    Forces were randomly sampled from this test set. 
    The total test energy and forces RMSEs are displayed at the top of this plot. 
    \textbf{(b)} Plot of the difference in predictions between NNP and DFT forces against the reference DFT forces. 
    }
\end{figure*}

\newpage

\begin{figure*}[p!]
    \centering
    \includegraphics[scale=0.78]{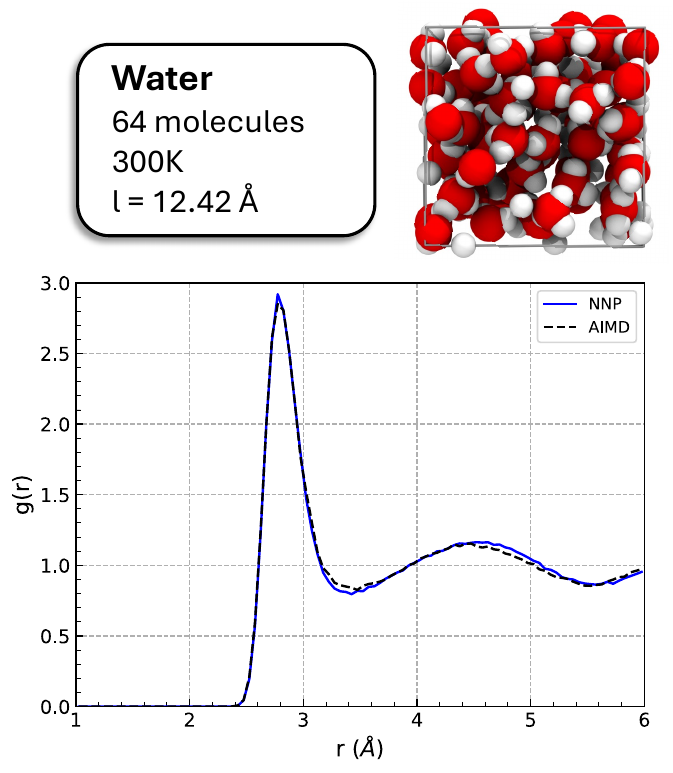}
    \caption{\label{fig:h2o_val}
    Radial distribution functions, $g(r)$, obtained for pure water at 300 K.
    Results are shown for both \textit{ab initio} and NNP MD. 
    Simulations were performed under the $NVT$ ensemble, with fixed numbers of particles $N$ (64 waters), fixed volume $V$ ($l_x=l_y=l_z=12.42$ \AA{}), and fixed temperature $T$ (300 K). 
    In total, 21 ps worth of \textit{ab initio} data and 1 ns worth of NNP data were obtained for RDF calculations.
    }

    \vspace{5mm}
        
    \includegraphics[scale=0.78]{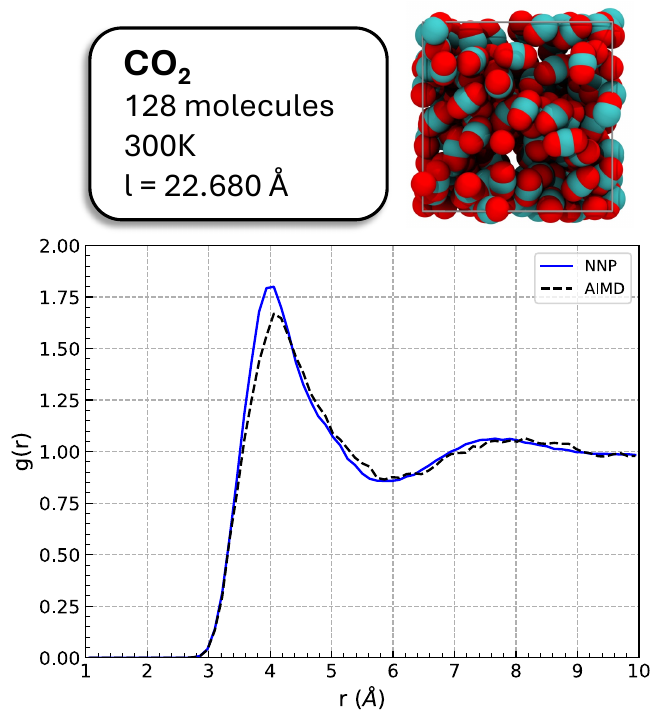}
    \caption{\label{fig:co2_val}
    Radial distribution functions, $g(r)$, for pure \ch{CO2} at 300 K. 
    Results are shown for both \textit{ab initio} and NNP MD. 
    Simulations were performed under the $NVT$ ensemble, with fixed numbers of particles $N$ (128 \ch{CO2}s), fixed volume $V$ ($l_x=l_y=l_z=22.680$ \AA{}), and fixed temperature $T$ (300 K). 
    In total, 6 ps worth of \textit{ab initio} data and 1 ns worth of NNP data were obtained for RDF calculations.
    }

\end{figure*}

\begin{figure*}[p!]
    \centering
    \includegraphics[scale=0.75]{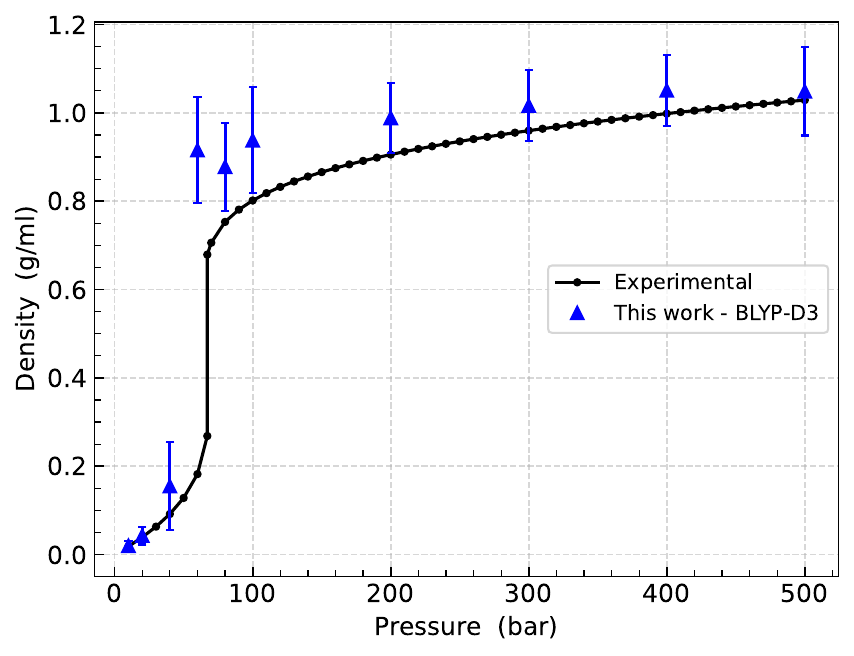}
    \caption{\label{fig:co2_dens}
    Densities predicted by NNP-MD for pure \ch{CO2} across 10-500 bar at room temperature. 
    Results are shown for both experiment (black, NIST WebBook) and from NNP-MD.
    Simulations were performed under the $NPT$ ensemble, with fixed numbers of particles $N$ (128 \ch{CO2}s), fixed pressure $P$, and fixed temperature $T$ (300 K). 
    System size was allow to vary isotropically with pressure (initial size, $l_x=l_y=l_z=22.680$ \AA{}).
    In total, 20 ns worth of NNP data was obtained for density calculations. 
    (Exp.: Roland Span and Wolfgang Wagner, \textit{J. Phys. Chem. Ref. Data}, 1996; \textbf{25} (6), 1509–1596).
    }
\end{figure*}

\clearpage

\newpage

\section{Size convergence}

\vspace{20mm}

\begin{figure*}[h]
    \centering
    \includegraphics[scale=0.62]{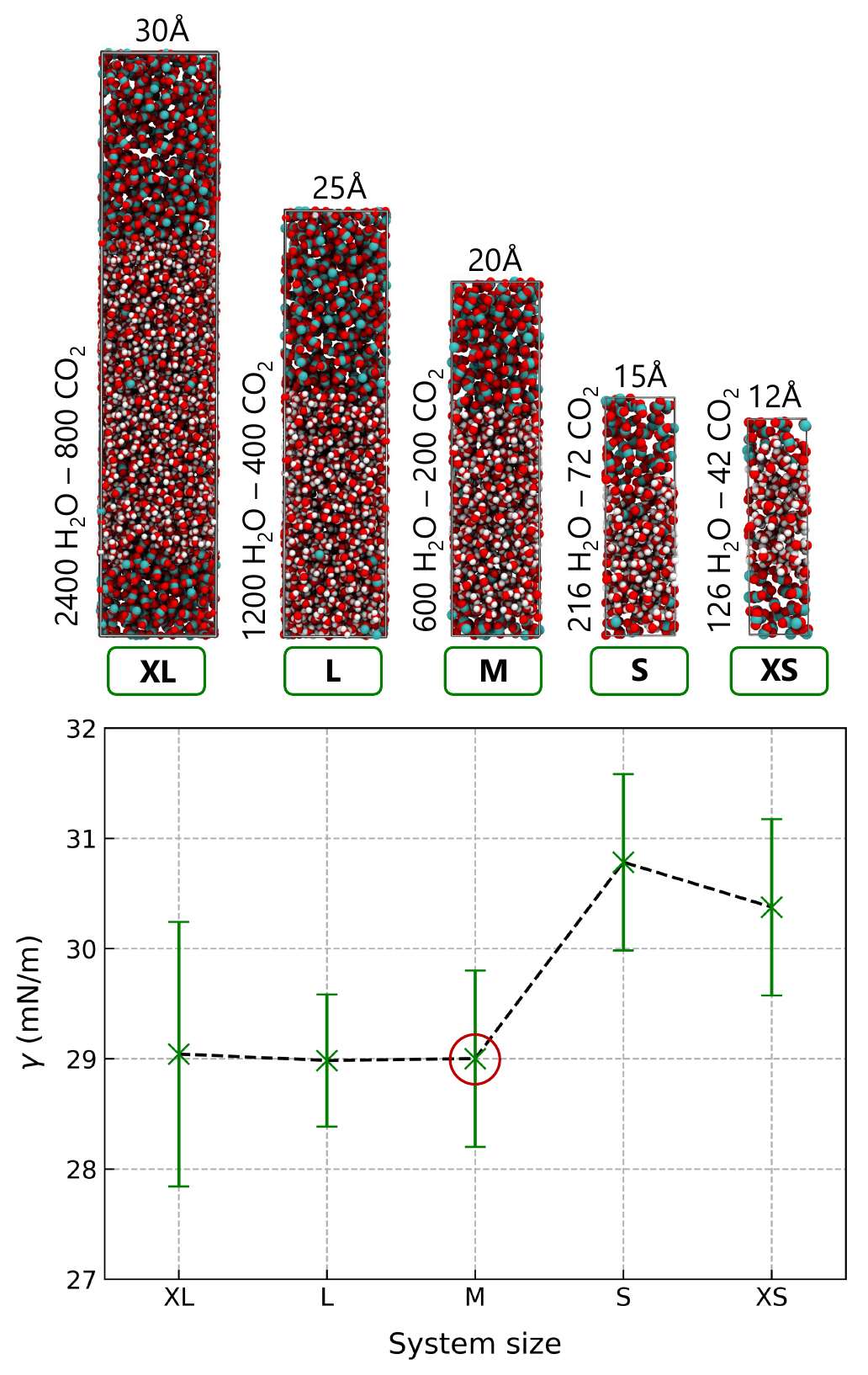}
    \caption{\label{fig:size}
    The effect of system size on the convergence of interfacial tension. 
    Plot shows the interfacial tension of \ch{CO2}-\ch{H2O} against system size for 300\,K and 30 MPa. 
    The setup selected for our production runs is highlighted by the red circle (M). 
    Representative snapshots of each of these systems are shown along the top of the figure. 
    Simulations were performed using SPC/E + EPM2 for 120 ns. 
    }
\end{figure*}

\newpage

\section{IFT Comparisons}

\vspace{50mm}

\begin{figure*}[h]
    \centering
    \includegraphics[scale=0.95]{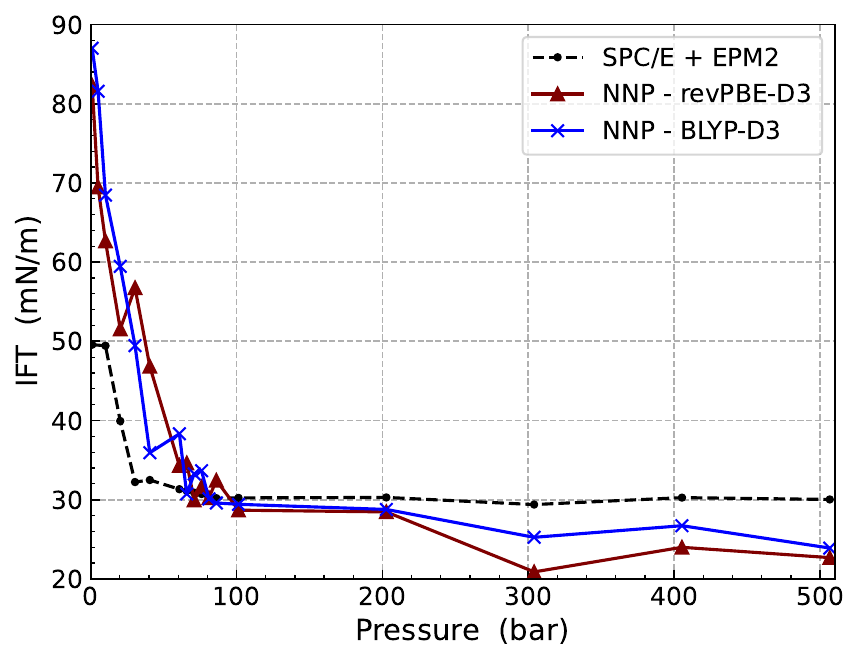}
    \caption{\label{fig:class_ift}
    Comparing interfacial tension profiles for NNP and classical results. 
    Interfacial tension profiles are shown for SPC/E + EPM2 (black), BLYP-D3 (blue) and revPBE-D3 (maroon, reduced dataset). 
    Classical runs were performed for 25 ns using the same system setup as with NNP-MD but without requiring replica exchange.
    }
\end{figure*}


